\documentclass[11pt,a4paper,usenames,dvipsnames]{article}

\usepackage[normalem]{ulem}
\usepackage[swedish,english]{babel}

\usepackage{color}
\usepackage{xcolor}
\usepackage{amssymb}
\usepackage{amsmath}
\usepackage{ae}
\usepackage{slashed}
\usepackage{graphics}
\usepackage{graphicx}
\usepackage{url}
\usepackage{hyperref}
\hypersetup{linktocpage}

\usepackage{cite}
\usepackage{times}
\usepackage{xfrac}

\usepackage{fancyhdr}
\usepackage{enumerate}
\def\be{\begin{equation}}
\def\ee{\end{equation}}

\usepackage{tikz}
\numberwithin{equation}{section}
\usepackage{tcolorbox}
\usepackage{enumitem}

\begin{document}
\selectlanguage{english}
\frenchspacing
\pagenumbering{roman}
\begin{center}
\null\vspace{\stretch{1}}

{ \Large {\bf
A scenario for AdS\textsubscript{2} and dS\textsubscript{2} emergence\\ from quantum fluctuations
}}
\\
\vspace{1cm}
Anna Karlsson
\vspace{1cm}

{\it Institute for Theoretical Physics, Utrecht University,\\
Princetonplein 5, 3584 CC Utrecht, The Netherlands}
\vspace{0.5cm}

{\it Department of Mathematical Sciences, Chalmers University of Technology\\ 
and the University of Gothenburg, SE-412 96 Gothenburg, Sweden}

\vspace{1.6cm}
\end{center}

\begin{abstract}
Spacetime emergence from entanglement has long been presumed to be holographic, with emergence from quantum interactions in a boundary theory. Since that setup is difficult to extend to general spacetimes, it is imperative to understand the key aspects of the map from entanglement to spacetime. We consider a scenario for spacetime emergence from quantum interactions directly in the spacetime, where the effective spacetime theory emerges from interactions similar to how fluid dynamics arises from molecule interactions. We show that this ansatz constitutes a working scenario for spacetime emergence in $2D$, with explicit examples for AdS and dS spacetimes. Importantly, by basing the emergence on interactions between quantum fluctuations that are present in the spacetime, key aspects of the dynamics observed in gauge/gravity duality analyses can be retained, at the same time as the quantisation can be applied to general spacetimes.
\end{abstract}

\vspace{\stretch{3}}
\thispagestyle{empty}
\newpage
\tableofcontents
\noindent\hrulefill
\pagenumbering{arabic}

\section{Introduction}
An emergence of spacetime from entanglement \cite{VanRaamsdonk:2010pw} between entities in a many-body system is a scenario supported by the gauge/gravity duality \cite{Maldacena:1997re,Witten:1998qj,Gubser:1998bc}. The typical many-body system in question is a set of particles in a gauge theory on the boundary of the Anti-de-Sitter (AdS) space, with a single spatial dimension emerging from the entanglement. In $2D$, a key model is the Sachdev--Ye--Kitaev (SYK) model \cite{Sachdev:1992fk,Kitaevx}. However, spacetimes with a boundary suitable for this kind of formulation constitute special cases. In attempts to model emergences of general spacetimes --- with a focus on de Sitter (dS) space --- increasing effort has been put into understanding precisely how the spatial bulk dimension (and gravity with it) emerges from entanglement. The key question is how to do away with the AdS boundary, while retaining a spacetime emergence from entanglement.

A possible and so far largely disregarded scenario for how entanglement (by quantum interactions) between particles may cause an emergent spacetime is in the same manner fluid dynamics arises from interactions between molecules in the fluid. A spacetime emergent from entanglement might simply be an effective theory governing a medium made up of interacting quantum particles, where the particles are present in the medium itself. A decided advantage of scenarios where spacetime emerges from degrees of freedom directly in the bulk, including the suggestion in \cite{Mertens:2022ujr}, is the adaptability to different types of spacetimes. A further argument for a fluid/molecule scenario is that any origin to the spacetime reasonably should have an interpretation relative to the spacetime itself, although such an interpretation need not be as local as in a fluid/molecule scenario. Suitable particles are however readily present in the vacuum: the quantum fluctuations.

We argue that a fluid/molecule scenario for spacetime emergence is realistic, provides interesting possibilities with respect to models of general spacetimes, and deserves further consideration. Following up on work in \cite{Karlsson:2019avt,Karlsson:2021ffs}, we describe how an effective theory of spacetime can arise from particle interactions governed by a set of particle-specific parameters, when the particles equivalently exist at a single point, for the cases of AdS\textsubscript{2} and dS\textsubscript{2} with no matter present. We also demonstrate how the spatial bulk dimension emerges under our ansatz when the time is kept continuous, and how the same type of emergence applies to both AdS and dS. These examples provide an interesting contrast to the mechanisms behind the emergence of the radial space direction in the gauge/gravity duality. Our hope is that our alternative way of approaching spacetime emergence from entanglement might facilitate determining the essential properties of the map from entanglement to emergent spacetime; what is crucial, and what is model-specific.

Although the spacetime emergence in the ansatz is fundamentally different from that of the gauge/gravity duality, when analysing the properties of the construction we find parallels to features observed in gauge/gravity duality analyses. In particular, an interconnecting mesh of particles similar to molecules in a fluid would have certain similarities to tensor networks \cite{Swingle:2009bg,Swingle:2012wq} while still decidedly different. In a fluid/molecule scenario, the bulk would be filled with interconnected particles, much like the tensor network nodes, but their positions and interactions would not be limited to a lattice. Instead, their interactions would be probabilistic, and local only in the effective theory (i.e. not at `small scales'). Their distribution would also be probabilistic, and the particles would be allowed to move. In fact, each particle must behave just like any particle in the spacetime, and follow a geodesic. This in turn sets the particle propagator to (in the effective theory) exactly coincide with the spacetime one. In addition, under the suggested ansatz it is not necessary to go to the boundary to describe the particle interactions --- while there is a similarity in distribution and interconnection to tensor networks, the theory in no way extends from the boundary. Any small patch of spacetime could be represented with a sufficiently large set of interacting particles.

Strange though such an ansatz may seem, one can argue that steps already have been taken in a similar direction, with analyses of SYK showing lattice field theories with a dynamical lattice length \cite{Lin:2022rbf}, allowed to fluctuate, and where chord diagrams introduced to keep track of which particles interact show a preference for geodesic slices of the bulk. The radial space direction in AdS has also been argued to emerge from spins interacting with immediate neighbours on a `hopping' lattice \cite{Guo:2022and}, and other related suggestions include subexponential violations of locality \cite{Akers:2022qdl}. Considering that the boundary is not a special subspace in the suggested ansatz, there might also be points of contact with the ongoing work on generalised entanglement wedges \cite{Bousso:2022hlz}.

More interestingly though, a spacetime emergence from the average behaviour of a number of particles would be characterised by error correcting features, as have been observed for AdS space \cite{Almheiri:2014lwa}. The average giving rise to a unit spacetime would have to be robust, requiring a higher number of particles than strictly necessary to capture the behaviour of the spacetime unit, and under a mapping to e.g. the boundary (basically a removal of the radial direction) the particles would be distributed along the boundary, so that information from different segments of the boundary could be used to decide the behaviour of a given unit of spacetime.

In short, the suggested ansatz of quantising spacetime through letting it be an effective theory emergent from quantum interactions between quantum fluctuations constitutes {\it a} quantisation of spacetime, applicable to general spacetimes and with characteristics compatible with features observed in gauge/gravity duality analyses. It could constitute a step from a holographic emergence of spacetime to simply an emergence from entanglement (not by way of a boundary), with increased model freedom. In $2D$ the dynamics of the particles is particularly simple, providing the simplest setup for the emergence of the radial space direction, and that is the special case we focus on in this text. The most unique feature of the ansatz is that the particles the spacetime emerges from move along geodesics in the spacetime. It is natural that any particle that exists in the spacetime does this, but it also makes the ansatz quite different from previous approaches to spacetime emergence. For example, it is not immediately clear how to draw a parallel between that property and some feature of the SYK model. Importantly, enough key spacetime characteristics observed in gauge/gravity duality analyses are reproduced (in this first analysis) to warrant an interest in this ansatz for spacetime emergence. With the end goal of determining how spacetime emerges from entanglement, it is relevant to understand the role a fluid/molecule scenario might play for the emergence.

\subsubsection*{Summary of the ansatz}
The basics of the ansatz setup were introduced in \cite{Karlsson:2019avt,Karlsson:2021ffs}. In summary, it is as follows. Consider a large set of particles, each with interactions in $d$ quantum channels. The possible quantum states of each of the particles are then described by an $S^{d-1}$ in spacetime. Observe that it is equally possible to define the presence of an $S^{d-1}$ in spacetime by the quantum states such a particle can assume. For example, the $\pm$ states of a qubit can either be defined by that the particle assumes two states when measured in the spacetime, or the presence of a spatial direction can be defined by that the $S^0$ particle assumes two distinct states (up/down).

Next, posit that each particle is endowed with a set of parameters, has a probability of being created that depends on said set of parameters, and that each pair of particles $(i,j)$ interact with a probability $P(i\leftrightarrow j)$ that depends on the sets of parameters of the two particles. Provided a suitable \mbox{$P(i\leftrightarrow j)$}, this probability of interaction can define spatial length; the distance between two particles can equivalently be set by their rate of interaction. For a set of particles interacting in $d$ quantum channels, initially not positioned in a spacetime, this construction leaves us with an interconnecting mesh in $d$ dimensions.

Finally, let each particle initiate interaction and changes in the particle parameters, order an event log in terms of consecutive events, and let time be defined by said event log and by that each particle interacts with a particle-specific frequency, in the same way atomic clocks can be used to define time. While the present discussion concerns $(d+1)$-dimensions, this type of construction can easily be extended to general dimensions.

With a suitable choice of particle parameters and associated interactions $P(i\leftrightarrow j)$, this ansatz can be used to model spacetimes, as exemplified in \S\ref{s.sAdS}-\ref{s.dS}. In the ansatz, the particles equivalently exist at a single point, but with an averaging over the statistical particle interactions, an effective spacetime theory can be obtained. With respect to that theory, the degree of entanglement between the particles corresponds to the distance between them (less interaction equals greater distance), multiple directions arise from a presence of interactions in multiple quantum channels, and time roughly gets converted into an event log logging a series of particle interactions and changes in parameter values.

From a top-down perspective, described in more detail in \S\ref{s.top}, the ansatz roughly corresponds to distributing a set of basis functions in the spacetime, with an even distribution on the surface when embedded in Minkowski space. However, instead of a tensor network extending from the boundary, the basis functions roughly fill the role of molecules in a fluid. They and their interactions both make up the effective, large-scale theory, and move in that same effective medium (the spacetime), same as any other particle. That is, they follow geodesics.

\subsubsection*{Summary of observations}
The ansatz describes a dynamic mesh of entangled, interacting quantum particles, all quantum fluctuations. We do not consider particles out-of-the vacuum (e.g. photons or matter) in the present text. The point is that given the effective medium, any particle in it (i.e. interacting with the set of particles) effectively behaves as though it moves in a spacetime, and a particle out-of-the-vacuum could interact with that medium and behave as a particle moving though a modelled spacetime.

The suggested ansatz for spacetime emergence really works best if all the spacetime dimensions are deconstructed simultaneously, but one can apply the emergence to e.g. one spatial dimension as well. We illustrate this in our examples. As such, AdS\textsubscript{2} can be described as a set of particles interacting in a theory with a time dimension only. The resulting model is not necessarily on the AdS boundary, but can be put there if one so wishes. The model fashioned in this way describes a large set of particles characterised by a parameter $b$ which describes the particle interaction frequency through $f=\sqrt{\alpha^2+b^2}$. This frequency appears to connect to the same quantity that in a gauge dual would specify the inverse temperature $\beta$. We observe that this is in line with the UV/IR correspondence: high-energy features in the gauge dual correspond to low-energy features in the gravity theory, and vice versa. The probability for particle creation $\propto f^{-1}$. In addition, the particle interaction rate decreases with disparity in $b$, and there is a spontaneous, uniform increase in entanglement between `nearby' particles. In comparison, dS\textsubscript{2} reduced in the same way is characterised by a parameter on an $S^1$ without any variation in interaction frequency, and the particles spontaneously get less entangled with time. For more details, please look at the respective sections in the main text. An interesting aspect is that this type of reduction of a spatial dimension may give relevant clues as to how spacetimes other than AdS need to be represented, in terms of how they emerge from entanglement. In the comparison between AdS\textsubscript{2} and dS\textsubscript{2}, a key difference is precisely the difference in parametric representation: a parameter connecting to energy vs. one taking values on an $S^1$, with no impact on the frequency of interaction.

Overall, in AdS\textsubscript{2} the mesh of interacting particles in the ansatz at any given moment in time would appear a bit similar to a tensor network, in distribution and interconnection. However, the particles in the ansatz do not conform to a lattice, the interactions are probabilisitic (not restricted to nearest neighbours or a lattice) and with time the parameters would change in accordance with the particles moving along geodesics in the spacetime. For a particular smallest region of spacetime to `form', a sufficient number of particles would need to interact, but the resulting unit spacetime could be in any region of the spacetime, as set by the parameters of the particles; the spacetime does not need to be built from the boundary. In addition, assuming such a unit spacetime to be robust in the sense of changing little with the addition or subtraction of a particle, one can consider a scenario where a unit spacetime is characterised by that exactly more than half of the particles are required to get an accurate prediction of the characteristics of the spacetime element. As we argue in \S\ref{s.error}, this would come with error correcting features with respect to the theory where the radial space direction has been removed. Under a removal of the radial space direction in AdS\textsubscript{3}, spacetime elements would be emergent from particles dispersed along an $S^1$ in a way that appears compatible with the error correcting features observed for AdS.

There are a few model choices that have been kept open in the present text; two of them more interesting. The first is whether the degree of freedom giving rise to a non-zero Ricci scalar corresponds to a fixed parameter, or a dynamic one subject to equilibration among interacting particles. This is relevant for spacetimes where the curvature is not constant. Under the present model ansatz, the Ricci scalar governs how a particle spontaneously in-/decreases its entanglement with other particles. If this is not a dynamic parameter, $2D$ spacetimes would be fairly rigid and uninteresting. In addition, we start with an initial assumption that the particle distribution is proportional to the spacetime volume element, but observe that for dS this is not required, which introduces a few interesting scenarios. We discuss those further in \S\ref{s.dS}.

\subsubsection*{Relation to quantisations of higher-dimensional spacetimes}\label{s.hD}
In the ansatz, the spacetime emerges from particles, their interactions and their dynamics. The particle dynamics is characterised by different types of degrees of freedom that correspond to specific parts of the Riemann tensor. $2D$ is especially simple, since the Riemann tensor is uniquely determined by the Ricci scalar $R$. Then there is only one degree of freedom to the dynamics: $R$ specifies the rate at which a particle spontaneously in-/decreases its entanglement with other particles. With a constant $R$, the behaviour is uniform throughout the spacetime. We analyse the associated dynamics in \S\ref{s.sAdS}.

A deconstruction of a general spacetime could be carried out in a manner similar to our deconstructions of AdS\textsubscript{2} and dS\textsubscript{2}, but would require more degrees of freedom with respect to the particle dynamics. The positioning of the particles in a target spacetime is in itself trivial: any metric $g_{\mu\nu}(x^\sigma)$ can be regarded as the product of a number of particles, each endowed with a set of parameters $\{x_i^\sigma\}$ and Gaussian interaction ranges so that a diagonalisation of the matrix $g_{\mu\nu}(x_i^\sigma)$ corresponds to the expected standard deviations for the particle $i$ in terms of $(2\sigma^2_i)^{-1}$, exactly as in the construction in \S\ref{s.sAdS}. This is a decided advantage of the ansatz: there is no conceptual obstruction to deconstructing any spacetime in this manner.

However, to specify the dynamics of the particles in a general spacetime so that they equivalently follow geodesics in the effective spacetime is a non-trivial matter in terms of execution. It is certainly feasible (there is no practical obstruction to the particles following geodesics) but in general the Riemann tensor has parts corresponding not only to $R$, but also a traceless part of the Ricci tensor, and a part that does not contribute to the Ricci tensor. The traceless part connects to a presence of charged particles moving in the spacetime, giving the effect of an electromagnetic field, and the final set of degrees of freedom connects to a random walk dynamics \cite{Karlsson:2021ffs}. In addition, in general spacetimes the average momentum of each particle does not need to be zero.

Altogether, a deconstruction of a general spacetime is beyond the scope this text, and not necessary for our purposes. We focus on showing how the emergence of the radial direction in AdS and dS can be modelled, and what the key properties of the ansatz are, in the simple low-dimensional settings which currently are relevant in analyses of the emergence of AdS and dS spaces. However, it is important that an extension of the ansatz to other spacetimes is not a conceptual issue, for the reasons stated above. Also, note that we in \S\ref{s.error} discuss a deconstruction of AdS\textsubscript{3}, which generalises directly from the case of AdS\textsubscript{2} since the only dynamics present is by $R$. In $3D$ a traceless part of the Ricci tensor could also be present, but that is not the case for AdS\textsubscript{3}. Instead, $R_{\mu\nu}\propto R g_{\mu\nu}$.

\subsubsection*{Outline}
In \S\ref{s.top} we begin with giving an intuitive picture of how one can approach a deconstruction of spacetime into particle interactions in the manner of a fluid/molecule scenario for the emergence of the effective medium. In \S\ref{s.sAdS} we proceed with describing how the spatial direction in AdS\textsubscript{2} can be modelled from a set of parameters governing particle interactions, and in \S\ref{s.error} we observe how that setup can connect to error-correcting features of the spacetime. In \S\ref{s.AdS} we proceed with describing the full quantisation for AdS\textsubscript{2} under the present ansatz, and in \S\ref{s.dS} we do the same two-step deconstruction for dS\textsubscript{2}. The procedure itself is not much different in the two cases. Finally, in \S\ref{s.outlook} we comment on relevant improvements of the model as well as on some of the possibilities the ansatz opens up with respect to modelling spacetime and spacetime features.

\section{Top-down illustration of the spacetime quantisation}\label{s.top}
A rough picture of the suggested deconstruction of the spacetime as seen from the effective theory (i.e. the spacetime) is the following. While not accurate in every detail, this illustration is useful for understanding the principles behind the model ansatz.

The ansatz is to quantise spacetime through letting particles in it act as basis functions for the spacetime, with spacetime emerging only as an effective theory (at large scales). From a top-down perspective this means that a large number of basis functions are distributed in the spacetime.

A standard way to discretise a smooth function is by Gaussian basis functions. We consider Gaussian basis functions partly (but not only) for the same reason; they constitute the most suitable choice as discussed in \cite{Karlsson:2019avt,Karlsson:2021ffs}. Also note that the position and momentum of each quantum particle are conjugate variables, and Gaussian distributions for the spacetime position and momentum correspond to the minimal uncertainty required by Heisenberg's uncertainty principle. A typical discretisation of a smooth function is given by
\be\label{eq.1dd}
f(x)=1\,,\,x\in\mathbb{R}\quad\rightarrow\quad f_d(x)=L\sum_{j\in \mathbb{Z}} \frac{1}{\sigma_j\sqrt{2\pi}}e^{-\frac{(x-jL)^2}{2\sigma_j^2}}\,,\quad \sigma_j=const\,,
\ee
where $L$ is a lattice spacing between the basis functions and $\sigma_j^2$ is the variance of each Gaussian. In the above, $\sigma_j$ defines a length scale, and the discretised function is characterised by
\be
\lim_{L/\sigma_j\rightarrow0}f_d(x)=f(x)\,.
\ee

In the spirit of the example above, consider a quantisation of spacetime in the following sense. Distribute a set of $N$ points $\{x^\sigma_j\}$ evenly on the $D$-dimensional surface embedded in Minkowski space and characterised by $ds^2=g_{\mu\nu}dx^\mu dx^\nu$. Next, assign a basis function to each point $x_j^\sigma$ in the form of a $D$-dimensional multivariate Gaussian distribution --- an $S^{(D-1)}$ with a Gaussian fall-off in the radial direction --- relative to the reference frame where $g_{\mu\nu}(x_j^\sigma)=\eta_{\mu\nu}$. We will refer to this particular reference frame as the particle rest frame. In that local frame, let each particle be characterised by the same variance, $\sigma_j^2=(2\pi a^2)^{-1}$, where $a$ sets a length scale (let $c=1$). The basis function at $x_j^\sigma$ is then given by
\be
a^De^{-a^2\pi \delta_{\mu\nu}(x-x_j)^\mu(x-x_j)^\nu}
\ee
in the particle rest frame. This expression represents a linearisation of the $ds^2$ at the point $x_j^\sigma$, and can be made independent of the reference frame employed (and diffeomorphism invariant) through a replacement of the $\delta_{\mu\nu}$ by $(g_E(x_j^\sigma))_{\mu\nu}$, where $(g_E)_{\mu\nu}$ is the metric in Euclidean signature. In a general reference frame, the basis function at $x_j^\sigma$ is given by
\be
a^D\sqrt{|g(x_j^\sigma)|}e^{-a^2\pi (g_E(x_j^\sigma))_{\mu\nu}(x-x_j)^\mu(x-x_j)^\nu}\,,
\ee
where $g$ is the determinant of the metric.

A discretisation of the spacetime performed in this way leads to that the continuous medium (spacetime) is replaced by a sum over probability distributions for the positions of a number of ($N$) particles in the spacetime,
\begin{gather}
\begin{aligned}\label{eq.tdd}
&\sum_j P(x^\sigma|j)=\sum_j a^D\sqrt{|g(x_j^\sigma)|}e^{-a^2\pi (g_E(x_j^\sigma))_{\mu\nu}(x-x_j)^\mu(x-x_j)^\nu}\\
&=\int dy^D \rho(y^\sigma)a^D\sqrt{|g(y^\sigma)|}e^{-a^2\pi (g_E(y^\sigma))_{\mu\nu}(x-y)^\mu(x-y)^\nu}\,,
\end{aligned}
\end{gather}
where the density function for the particle distribution is
\be\label{eq.rho}
\rho(y^\sigma)=\sum_j \delta(y^\sigma-x^\sigma_j)\,.
\ee
Moreover, the even distribution of the basis functions in the spacetime when embedded in Minkowski space (where spacetime distance is accurately portrayed) means that the probability of finding a particle centred at $x_j^\sigma$ is proportional to the spacetime volume element $\sqrt{|g|}$ as well as the number ($N_{u}$) of particles distributed per unit area. We therefore have
\be\label{eq.rho}
\rho(y^\sigma)= N_u\sqrt{|g(y^\sigma)|}
\ee
relative to the effective theory.

In total, \eqref{eq.tdd} basically is a multivariate counterpart of \eqref{eq.1dd}, except for that the expression has been made diffeomorphism invariant through $g_{\mu\nu}(x_j^\sigma)$. As a consequence, variations in length scale across the surface are also incorporated. In addition, $\rho(y^\sigma)$ allows for a random distribution of the basis functions (not restricted to a lattice) and the expression is not normalised: the overall factor $L$ in \eqref{eq.1dd} corresponds to dividing \eqref{eq.tdd} by $N_u$.

In essentials, the suggested quantisation corresponds to exchanging the notion of a continuous spacetime in which everything is situated for a medium upheld by particles in the vacuum, and their interactions. A particle's position in the spacetime is exchanged for a notion of which particles it (mostly) interacts with. Local notions of direction and length get substituted with interaction rates in a set of quantum channels. From a top-down perspective the suggested ansatz amounts to replacing the spacetime with a set of overlapping probability distributions, given by \eqref{eq.tdd}. In that way, the continuous spacetime (in the present ansatz interpreted as an effective medium) is rephrased in terms of a set of particles and their quantum interactions. A key point is that a continuous spacetime medium {\it can} be replaced with, and upheld by, overlapping probability distributions in this way. From a large-scale perspective, the described medium would appear to be continuous.
\\\\
Note that while the description above captures several key features of the model ansatz, it falls short of giving a full picture of it. To begin with, a top-down analysis of an effective theory never captures the statistical fluctuations that are allowed around the mean properties present in the effective theory. The $g_{\mu\nu}(x_j^\sigma)$ in \eqref{eq.tdd} is such a mean property. In general, it needs to be replaced by a $(\mathtt{g}_j)_{\mu\nu}$ with
\be\label{eq.met}
\langle\mathtt{g}_{\mu\nu}\rangle_{dV}=g_{\mu\nu}
\ee
when the average is taken over a spacetime volume element $dV$. At the particle level, $g_{\mu\nu}$ represents a mean over several interacting particles regarding the delay in communication between them in different quantum channels (distance in different directions) relative to an event log (time as specified by atomic clocks). The number must be large enough for perturbations in $g_{\mu\nu}$ to be small with respect to individual particles.

Moreover, the presented quantisation represents a naive snapshot in time: there has been no mention of time dependence. Each particle has a lifetime and will propagate through the spacetime, and the $P(x^\sigma|i)$ of the basis functions in \eqref{eq.tdd} must reflect that propagation. The $P(x_j^\sigma)$ cannot be set solely by the random distribution in \eqref{eq.rho}.

The full picture of what the ansatz entails is best given in a bottom-up description, which we will proceed with shortly. However, a few more key features can be identified from the quantisation scheme outlined above. To begin with, there is a singe unidentified scale $a$ in \eqref{eq.tdd}. $(2\pi a^2)^{-1}$ corresponds to the variance of the position probability distribution of each particle in the spacetime, i.e. a minimal uncertainty distribution due to the quantum nature of the particle. As always, an uncertainty in spacetime position comes with an uncertainty in momentum, and Heisenberg's uncertainty principle ($2\sigma_x\sigma_{p_x}\geq\hbar$) implies the distributions to be characterised by $\sqrt{2}\sigma_x=l_p$ and $\sqrt{2}\sigma_{p_x}=\hbar/l_p$ in an equilibrium state (not perturbed by measurements), with
\be\label{eq.a}
(\pi a^2)^{-1}=l_p^2\,.
\ee

From the quantisation, it also follows that the probability for two particles to interact is given by the overlap of the independent position probability distributions of the particles; from the spacetime perspective, two particles interact if they are at the same position in the spacetime. Given that two particles $(i,j)$ exist, the probability for them to interact is
\be\label{eq.tddi}
P(i\leftrightarrow j|i,j)=\int dx^DP(x^\sigma|i)P(x^\sigma|j)=a^D\sqrt{\mathtt{g}_{ij}}e^{-a^2\pi (\mathtt{g}_{ij})_{\mu\nu}(x_i-x_j)^\mu(x_i-x_j)^\nu}\,,
\ee
where 
\be
(\mathtt{g}_{ij})_{\mu\nu}=(\mathtt{g}_{i,E})_{\mu\rho}(\mathtt{g}_{i,E}+\mathtt{g}_{j,E})^{\rho\sigma}(\mathtt{g}_{j,E})_{\sigma\nu}\,.
\ee
Here, $(\mathtt{g}_{i,E})_{\mu\nu}$ is just a matrix: $(\mathtt{g}_i)_{\mu\nu}$ in Euclidean signature. Observe that $P(i\leftrightarrow j|i,j)$ is very well-behaved: it has a Gaussian fall-off with the distance between the particles.

From the Gaussian fall-off of $P(i\leftrightarrow j|i,j)$, as well as from \eqref{eq.met}, it is also clear that the number of particles ($N_u$) per volume element in the spacetime must be very large in the suggested model. An interconnecting quantisation of this kind requires that the particles interact frequently for an effective, continuous medium to emerge from the particles and their interactions.

\section{Removing the spatial dimension in AdS\textsubscript{2}}\label{s.sAdS}
To connect the model ansatz to the present approaches to spacetime emergence --- or at least to show what the ansatz looks like in that setting --- we begin by describing how the model ansatz can be employed to remove the spatial direction in AdS\textsubscript{2}. First, consider the metric as given by
\be\label{eq.AdS2m}
ds_\text{AdS\textsubscript{2}}^2
=\frac{-d t^2+dx^2}{\cosh^2([t-t_o]/\alpha)}
=-(\alpha^2+b^2)d\tilde t^2+\frac{db^2}{1+b^2/\alpha^2}
\,,\quad \text{with}\quad R=-2/\alpha\,.
\ee
These two different sets of coordinates are useful in that the $(t,x)$ coordinates can be set to coincide with the particle rest frame at any given spacetime point $x_o^\sigma$, and the $(\tilde t,b)$ are a simple rephrasing of global coordinates with the spatial dimension vanishing in the limit where $|b|\rightarrow\infty$. In the present ansatz, the latter set of coordinates gives a useful set of parameters for the particles that give rise to the spacetime, and we quantise the spatial direction with respect to $b$.

Following the general outline of the quantisation described in the previous section, consider a setting where particles characterised by a parameter $b$ are generated with a probability density\footnote{Note that a key property of the ansatz is that the interaction ranges \eqref{eq.bdist} in the quantised theory specify distances in the effective theory. For this reason they must correspond directly to the target spacetime metric; a construction which coincides with the uniform behaviour in \eqref{eq.reqA2}. The density is only relevant insofar as there must be sufficient interaction for the effective theory to form. We treat our initial assumption of $\rho\propto\sqrt{|g|}$ as an open choice, and discuss possible options for $\rho$ in \S\ref{s.dS}. For AdS, repulsive particle interactions and time invariance set the density to be that of \eqref{eq.rhob}.}
\be\label{eq.rhob}
\rho_b( b)\propto(1+b^2/\alpha^2)^{-1/2}\,,\quad b\in\mathbb{R}\,.
\ee
Let each particle generated in this way and characterised by a value $b_i$ have an uncertainty in the value of said parameter,
\be\label{eq.bdist}
P(b|i)= \frac{1}{\sigma_{bi}\sqrt{2\pi}} e^{-\frac{(b-b_i)^2}{2\sigma_{bi}^2}}\,, \quad \left\langle2\sigma_{bi}^2\right\rangle^{-1}=\pi a^2(1+b_i^2/\alpha^2)^{-1}\,,
\ee
so that $b_i$ is the expected value of $b$ for the particle $i$. In the effective (spacetime) interpretation, the scale $a$ is given by \eqref{eq.a}. Note that the $P(b|i)$ is equivalent to that the particle is endowed with an $S^0$ with the specified Gaussian fall-off in the radial direction.

Furthermore, let a temporal dimension exist, let $b_i=b_i(\tilde t)$ and let particles interact if they assume the same value of $b$ at the same time. As in \eqref{eq.tddi}, the interactions are then given by
\be\label{eq.bint}
P(i\leftrightarrow j|i,j)=\int db\,P(b|i)P(b|j)\,.
\ee

So far, we have followed the procedure described in the previous section. Proceeding with the time dependence of $b_i(\tilde t)$, let the particles interact with a frequency dependent on $b$, 
\be\label{eq.fAdS}
f(b)=\sqrt{\alpha^2+b^2}\,,
\ee
and introduce what corresponds to a momentum through letting $\langle \partial_{\tilde t} b_i \rangle=0$ at particle creation, with small variations around the mean. Here, the zero is just the mean momentum of each spacetime element in the chosen reference frame $(\tilde t,b)$. Note that these separate introductions of frequency and momentum are necessary only when time is kept continuous. In \S\ref{s.AdS} they arise from the model.

Furthermore, let the particles {\it spontaneously and uniformly} get more entangled with time. From a spacetime point of view, this corresponds to a spontaneous increase in particle interaction range with time\footnote{I.e. each particle gets more entangled with the particles in its environment, without a change in interaction rate.} so that the standard deviation of the position probability distribution increases with time. Effectively, this will make the entire spacetime shrink towards $b=0$, in accordance with that an increase in entanglement between particles corresponds to a decrease in distance between them, and $b_i(\tilde t)$ will be a function of time in precisely such a way that each particle follows a geodesic.

To show the effect of a uniform change in degree of entanglement, we formulate the increase in interaction range relative to a frame that coincides with the particle rest frame at a given point $x_o^\sigma$ in spacetime. Recall that $g_{\mu\nu}=\eta_{\mu\nu}$ in the particle rest frame. The main advantages of the particle rest frame is that the time in that frame gives the proper time of the particle, that the particle behaviour relative to said frame is constant in time, and (at least in our example of AdS, and later dS) each particle behaves identically relative to its particle rest frame, up to a particle-specific clock frequency (see \S\ref{s.AdS} for more details).

To capture the uniform change in degree of entanglement, we need a frame that coincides with the rest frame at $x_o^\sigma$ while allowing for a dynamic particle behaviour. Call that frame $(t,x)$. Let the $(t,x)$ frame be characterised by that the standard deviation $\sigma_i$ is the same for both coordinates and only dependent on $t$, and impose a uniform behaviour through a condition on $\partial_t^2\sigma_i(t)$:
\be\label{eq.reqA}
\sigma_i(t_o)=\frac{l_p}{\sqrt{2}}
\,,\qquad \partial_t\sigma_i(t)\big|_{t=t_o}=0\,,\qquad\frac{\partial_t^2\sigma_i(t)}{\sigma_i(t)}=\frac{1}{\alpha^2}\,.
\ee
\be\label{eq.reqA2}
\eqref{eq.reqA}\quad\Rightarrow \quad\sigma_i=\frac{l_p}{\sqrt{2}}\cosh\left(\frac{t-t_o}{\alpha}\right)\quad \Rightarrow\quad ds^2=\frac{-dt^2+dx^2}{\cosh^2([t-t_o]/\alpha)}\,.
\ee
This frame coincides with the particle rest frame at $t=t_o$, and the metric imposed by the uniform behaviour in \eqref{eq.reqA} is the same as for the $(t,x)$ coordinates in \eqref{eq.AdS2m}. Note that while we have constructed the requirements in \eqref{eq.reqA} so that a patch of AdS\textsubscript{2} is reproduced around $x_o^\sigma$, from a bottom-up perspective it is equivalent to impose a uniform, spontaneous increase in entanglement between a particle and its environment by \eqref{eq.reqA} to get the AdS\textsubscript{2} metric in the $(t,x)$ coordinates in \eqref{eq.AdS2m}. Also note that while the condition in \eqref{eq.reqA} is phrased as an exact condition, small variations around the value of $\alpha^{-2}$ can be allowed, provided that $\alpha^{-2}$ remains the expected value.

Now, \eqref{eq.reqA2} captures the behaviour of the particle at the point $x_o^\sigma$. At that point, the acceleration of the particle is given by the geodesic equation in AdS\textsubscript{2}. If desired, one can express the $(\tilde t,b)$ coordinates in terms of $(t,x)$ and recapture\footnote{For the coordinate embeddings in Minkowski space, see appendix \ref{s.emb}. Also note that $\frac{d^2x}{ds^2}=\frac{d^2t}{ds^2}=0$ at $x^\sigma=x^\sigma_o$.} the geodesic equation in the $(\tilde t,b)$ coordinates from the conditions in \eqref{eq.reqA}, but it is equivalent to simply note that the fact that the particle obeys the geodesic equation in AdS\textsubscript{2} at the point $x_o^\sigma$ in the $(t,x)$-frame infers that it also does so in the $(\tilde t,b)$-frame, by diffeomorphism invariance. Moreover, the point $x_o^\sigma$ is not a special point; the particle behaviour can be described in this way at any point in AdS\textsubscript{2}. Hence the acceleration at every point in the spacetime is given by the geodesic equation, and the particles follow geodesics. That is, the condition that the interaction range of every particle increases uniformly with the proper time of the particle, as specified in \eqref{eq.reqA}, infers that the particles (the basis elements for the spacetime) follow geodesics in AdS\textsubscript{2}.
\\\\
In summary, this quantisation of the spatial direction consists of a (very large) set of particles that are characterised by a parameter $b$. Since the particles interact with a frequency $f$ proportional to $\sqrt{\alpha^2+b^2}$, we can identify $b$ as contributing to the energy of the particle. Particles with a specific $b$ are created with the probability density \eqref{eq.rhob} which is proportional to $f^{-1}$, there is an intrinsic uncertainty in the parameter $b$ by \eqref{eq.bdist}, and there is a Gaussian fall-off in interaction with $\Delta b$, by \eqref{eq.bint} and \eqref{eq.tddi}. The setup specifies a set of particles interconnected by quantum interactions which, if the rates of interaction between the particles is rendered as distance between points, makes up a medium which at large scales coincides with AdS\textsubscript{2}. Moreover, the particles spontaneously and uniformly increase their interaction range with their proper time, by \eqref{eq.reqA}, and that causes an evolution of $b(\tilde t)$ which corresponds to the particle moving along geodesics in AdS\textsubscript{2}. Observe that the uniform shrinking of the AdS\textsubscript{2}-medium towards $b=0$ has no impact on the density distribution of the particles in the effective spacetime except in terms of a constant times the distribution in \eqref{eq.rhob}.

A feature of the ansatz is that what in a gauge theory on the AdS boundary figures as a period in the system time and therefore is interpreted as an inverse temperature $\beta$, in this suggested ansatz instead is proportional to the frequency $f$. The difference in interpretation is caused by the difference in basing the model on a periodicity in the boundary time vs. letting the curvature of the effective theory be created by a difference in frequency of interaction. This difference is in line with the UV/IR correspondence; under the ansatz, the quantisation retains the energy concepts of the gravity theory. This endows the quantisation with a new flexibility compared with what is present in a gauge dual. A particle's position in the spacetime can be interpreted in terms of the frequency $f$.

Moreover, note that in this quantisation, the spatial direction arises since each particle interaction is confined within a small range of $b$, and it takes time for information to propagate through the chain of particles obtained by ordering them according to their $b$s (by their energies, up to the sign of $b$).

\section{Relation to error correcting features}\label{s.error}
As visible in \S\ref{s.sAdS}, the suggested quantisation of the spatial direction in AdS\textsubscript{2} is quite different from the current standard approaches to spacetime emergence. In particular, the feature that the entanglement the spacetime emerges from constitutes entanglement between particles that move along geodesics in the spacetime is a decidedly distinct feature of the present ansatz. Interestingly though, the resulting model displays error correcting features, same as have been observed for AdS space \cite{Almheiri:2014lwa}. 

The $2D$ scenario that is the focus of this text is not ideal for analysing properties related to error correction --- the AdS boundary just consists of two points, and to reduce the model in \S\ref{s.sAdS} to the boundary (or rather, the corresponding concept in the ansatz) would simply correspond to separating the particles into two locations based on ${\rm sgn}(b)$ (with a random allocation for $b=0$) and letting $b$ set the energy of each particle. To get a more interesting comparison between the suggested quantisation and error correcting codes, consider AdS\textsubscript{3} with the metric
\begin{gather}
\begin{aligned}\label{eq.AdS3m}
ds_\text{AdS\textsubscript{3}}^2&=-(\alpha^2+b^2)d\tilde t^2+\frac{db^2}{1+b^2/\alpha^2}+b^2d\theta^2 \\
&=\alpha^2(-\cosh^2\rho \,d\tilde t^2+d\rho^2+\sinh^2\rho \,d\theta^2)
\,,\quad \text{with}\quad R=-6/\alpha\,,
\end{aligned}
\end{gather}
where the proper distance is expressed in global coordinates in the second line ($b=\alpha\sinh \rho$). The radial direction of this spacetime can be modelled to emerge in the same way as for \eqref{eq.AdS2m}, because the Riemann tensor associated with the spacetime is fully determined by $R$, as discussed in \S\ref{s.hD}. That is, we can quantise $b$ in \eqref{eq.AdS3m} in the same way as for AdS\textsubscript{2} in \S3. We then get a setting with a $2D$ theory consisting of a spatial direction described by $\theta\in S^1$, and a time direction. At each point $(\tilde t,\theta)$ in that theory, there is a number of particles characterised by a parameter $b\in\mathbb{R}^+$, generated by a probability density \eqref{eq.rhob} and an uncertainty in the parameter value by \eqref{eq.bdist}. The only necessary addition is a counterpart to \eqref{eq.bdist} specifying an uncertainty in $\theta$ so that a particle with parameter $b_i$ placed at $\theta_i$ has a probability to interact with particles of different $\theta$ by a Gaussian distribution in terms of $\theta-\theta_i$, with $\langle2\sigma_{\theta i}^2\rangle^{-1}=\pi a^2 b_i^2$. Interactions occur when particles assume the same values of $(\tilde t,b,\theta)$, etc.

Next, keep in mind that the emergence of the effective theory is analogous to how hydrodynamic theories emerge: when the large-scale limit is taken, the microscopic dynamics is averaged over until only mean, long-lived properties remain. In a fluid, a number of interacting particles give rise to a volume element of the fluid, and in our model ansatz a number of interacting particles give rise to a volume element in the spacetime.

If we consider a time slice of the spacetime, i.e. the particles present at a time $\tilde t=\tilde t_o$ in this quantisation of AdS\textsubscript{3}, the model has the property that a certain number of particles with approximate values ($\theta_o,b_o$) need to be present for the effective medium, spacetime, to emerge in terms of an area element around ($\theta_o,b_o$). Let each unit of emergent spacetime be a product of an average over the behaviour of $m$ particles, in a way so that more than half of those particles are needed to get an accurate representation of the unit spacetime element. An area element around a point ($\theta_o,b_o$) can be described by $db d\theta$, or equivalently by $d\rho d\theta$, and since the particle distribution is \mbox{$\propto\sqrt{|g|}$}, the number of particles $n$ in a small area element\footnote{Small enough for a linearisation around that point to be an accurate description of the particle density. We also assume that $b_o$ is not too small.} near the point ($\theta_o,b_o$), with values \mbox{$|\theta-\theta_o|\leq d\theta/2$} and $|\rho-\rho_o|\leq d\rho/2$ is characterised by
\be\label{eq.dens}
n\propto\alpha\sinh\rho_o\, d\theta\, d\rho=b_o\, d\theta\, d\rho\,.
\ee
Now, this rendition in terms of $d\rho$ makes it easier to identify the key features of the emergence. $d\rho$ captures the length element in the radial direction at any point in the spacetime, since the metric is diagonal with $g_{\rho\rho}=1$. What \eqref{eq.dens} then says is that for a constant $d\rho$, $d\theta$ must vary for a fixed number of $n$ particles to be present in the area element. In the model ansatz, $\theta$ describes a particle's position on an $S^1$, and $b=\alpha\sinh\rho$ is only a parameter assigned to each particle, so equivalently the portion of the $S^1$ that needs to be probed in order to represent an emergent area element in the spacetime depends on the position of that element in the spacetime. The smaller the value $b_o$ is (which the element is centred around) the larger the probed portion of the $S^1$ must be. A convenient way to visualise this might be to consider a small area element on a disk: the closer to the centre point the element lies, the wider the distribution of angular positions (in terms of polar coordinates) within that area element.

The key point here is how the particles that give rise to an emergent area element in the space are distributed on the $S^1$, in the lower-dimensional theory. In the setting described above --- including the condition that $m$ particles give rise to a unit element of spacetime, but only just over half are required to accurately represent the spacetime element --- error correcting features arise. For example, consider the area element emergent at $b_o=0$. To take all of the particles in that element into consideration, the full $S^1$ would need to be probed. However, to accurately represent the same element, only just over half of the $S^1$ would need to be probed, with respect to values $b\approx0$. The distribution on the $S^1$ is even, and the information is redundantly stored. To represent the area element any combination of segments of the $S^1$ could be probed, as long as the total add up to more than half of the $S^1$. Similarly, area elements centred around $b_o>0$ emerge from particles distributed along parts of the $S^1$, centred around $\theta_o$ and with increasingly smaller distributions in $\theta$ for larger values of $b_o$. Each such segment is characterised by error correcting features in the same way; it is sufficient to probe just over half of the associated particles to represent the emergent area element.

For a visualisation of how the space of the time slice is stored on the $S^1$ in the lower-dimensional theory (without the radial direction) consider the time slice of $AdS\textsubscript{3}$ as represented on the Poincar\'e disk. In comparison, any point in that space would be mapped to the $S^1$ based on the angular polar coordinate of its position on the disk, and unit elements closer to the centre would end up distributed along larger portions of the $S^1$. Moreover, note that through probing a segment of the $S^1$ indiscriminately with respect to $b$, one in this way would obtain information on a spacetime region very similar to an entanglement wedge in the bulk.
\\\\
Note that the redundant encoding of the spacetime in the ansatz does not extend across different spacetime regions, in the effective theory. As regarded from a spacetime perspective, the emergence is local, in the same way the emergence of a volume element in a fluid is local. However, in the quantised AdS\textsubscript{3} model, particles on different parts of the $S^1$ can interact, depending on the parameter value $b$.

Also, while the setup discussed above is a very special setting in that the radial direction can be quantised and the theory then described in terms of a $2D$ theory (a spatial $S^1$ and a time), the error correcting features of the ansatz extend to any other spacetimes that might be considered, in the sense described in \S\ref{s.hD}. That is, from the effective spacetime perspective, a volume element of the spacetime would emerge from a number of particles present locally, in said volume element, and not all of them would be needed to represent the volume element. In a quantised rendition of the theory, those same particles may be represented in ways which do not reflect their proximity in terms of how frequently they interact.
\\\\
In summary, it is interesting that the ansatz that spacetime emerges from an average behaviour of a set of interacting particles infers that the rendition of AdS spacetimes with one spatial direction quantised (corresponding to the parameter $b$ as illustrated above) is characterised by properties similar to those of error correcting codes, same as have been observed in gauge/gravity duality analyses. Under the present ansatz for quantisation, the error correcting features are a direct product of that a minimal numer of interacting particles is required to predict the behaviour of an emergent unit of the effective medium, spacetime. 

\section{A quantisation of AdS\textsubscript{2}, including time}\label{s.AdS}
In \S\ref{s.sAdS}, we showed how to break down the spatial direction in AdS\textsubscript{2} under the present ansatz for spacetime emergence, and we detailed how each particle under this ansatz propagates along a geodesic. When quantising time as well, the top-down quantisation is simply to introduce interaction points along the line the geodesic of the particle describes, at even intervals in the proper time of the particle, with a frequency corresponding to the clock frequency of the particle. This gives a series of interaction points along the geodesic, during the lifetime of the particle.

In a quantisation of both dimensions in AdS\textsubscript{2}, the spatial quantisation remains much the same as in \S\ref{s.sAdS}. Consider a setting where particles characterised by two parameters $(\tilde t, b)$ are generated with probability densities given by \eqref{eq.rhob} and
\be
\rho_{\tilde t}(\tilde t,b)\propto \sqrt{\alpha^2+b^2}\,,\quad\tilde t\in \mathbb{R}\,. 
\ee
Let each particle generated in this way and characterised by values $(\tilde t_i,b_i)$ have an uncertainty in the values of said parameters given by \eqref{eq.bdist} and
\be\label{eq.AdSpt}
P(\tilde t|i)=
\frac{1}{\sigma_{\tilde t_i}\sqrt{2\pi}} e^{-\frac{(\tilde t-\tilde t_i)^2}{2\sigma_{\tilde t_i}^2}}\,, \quad \left\langle2\sigma_{\tilde t_i}^2\right\rangle^{-1}=\pi a^2(\alpha^2 +b^2)\,,
\ee
with $P(\tilde t, b|i)=P(\tilde t|i)P(b|i)$. Here, $\tilde t_i$ is the expected value of $\tilde t$ for the particle $i$. Also, let particles interact if they assume the same values of $(\tilde t, b)$.

As observed in \S\ref{s.top}, each particle is equivalently characterised by an $S^{1}$ at each point of interaction, with a Gaussian fall-off in the radial direction with variance $(2\pi a^2)^{-1}$. Let each particle initiate interaction with a particle-specific rate in comparison to other particles (a clock frequency) so that the interaction points of a particle can be ordered and represented as
\be\label{eq.AdSts}
\tau_n=\tau_{n-1}+ s t_p\,, \quad n\in\left\{n:n\in\mathbb{N},n\leq \frac{\tau_l}{st_p}\right\}\,,
\ee
relative to the $\tilde t$-direction of the unit $S^1$ of the particle, at each step. Here, $\tau$ becomes the proper time of the particle, $s$ is the particle clock frequency\footnote{One can consider a set with different kinds of particles, where each type of particle is characterised by a specific clock frequency $s$. The present discussion is not at that level of detail.}, $t_p$ is the Planck time and $\tau_l$ is the lifetime of the particle. At particle creation the particle is characterised by $\tilde t_i$, and the next point of interaction is generated from the $S^1$ of the particle, giving a new interaction point $\tilde t_i+st_p$ relative to said $S^1$, and so on. At each of those interaction points, let the particle be characterised by an $S^{1}$ with a Gaussian fall-off in the radial direction with variance $(2\pi a^2)^{-1}$. This construction gives a series of interaction points (a series of ordered events) along a function $\tilde t_i=\tilde t_i(\tau)$. It also creates a frequency difference by $b$ as in \eqref{eq.fAdS}, since the unit distance in $\tilde t$ relative to the $S^1$ of a particle depends on $\sigma_{\tilde t_i}^2$ by \eqref{eq.AdSpt}. In addition, the alignment of the main axes of each $S^1$ will not be identical, and small variations in that alignment is what in a dimensional reduction of only the spatial dimension gives rise to an uncertainty in momentum, as described right after \eqref{eq.fAdS}.

Moreover, let the particles that do interact spontaneously get more entangled, i.e. let the particle interaction range relative to its unit $S^1$ increase, uniformly, at each new point of interaction. As shown in \S\ref{s.sAdS} this corresponds to that $b_i=b_i(\tau)$ so that the particle follows a geodesic.

In the above, we have given a description of what the interaction points of a particle mean relative to each particle, its parameters and its interaction interface, the $S^1$. It is a model with a series of events that can be ordered relative to the $S^1$ of the particle, where the particle becomes characterised by an ordering parameter $\tilde t$ (time) and another parameter ($b$) which can be interpreted in terms of energy. We have shown that the interactions between the particles create an interconnected medium which at large scales behaves as AdS\textsubscript{2}. Through their interactions, the particles make up the effective theory, and relative to it, they move around as any particle would while governed by the effective theory.

\section{Quantising dS\textsubscript{2}}\label{s.dS}
The quantising procedure described above can be applied to $D\leq3$ spacetimes where the Riemann tensor is uniquely specified by the Ricci scalar. A key difference between AdS and dS is that the latter cannot be reduced to a boundary theory where only time is present, but that is irrelevant under the present ansatz for spacetime emergence. For de Sitter space, a quantisation of only the spatial direction can be fashioned through using
\be\label{eq.dS2m}
ds^2_\text{dS\textsubscript{2}}
=\frac{-dt^2+dx^2}{\cos^2([t-t_o]/\alpha)}
=-d\tilde t^2+\alpha^2\cosh^2(\tilde t/\alpha)\,d\theta^2
\,,\quad \text{with}\quad R=2/\alpha\,.
\ee
These two different sets of coordinates are useful in that the $(t,x)$ coordinates can be set to coincide with the particle rest frame at any given spacetime point $x_o^\sigma$, and the $(\tilde t,\theta)$ are the global coordinates. In the present ansatz, the latter set of coordinates gives a useful set of parameters for the particles that give rise to the spacetime.

\subsubsection*{Quantising the spatial dimension}
A quantisation of only the spatial dimension can e.g. be done with respect to $\theta$. Consider a setting with a time dimension $\tilde t$ where particles characterised by a parameter $\theta$ which takes values on a unit $S^1$ are generated with a probability density
\be\label{eq.rhothe}
\rho_\theta(\tilde t,\theta)\propto\alpha\cosh([\tilde t- \zeta]/\alpha)\,,
\ee
where $\zeta$ is some constant that is set by $\tilde t$ and $\tilde t_l$. The need for some $\zeta$ is apparent: if the spacetime medium is supposed to have an even distribution of particles $\propto \sqrt{|g|}$ in it at the same time as $\rho_\theta$ changes with time and the particles have a non-zero lifetime, then there needs to be a shift in $\rho_\theta$, e.g. by $\zeta$. However, for dS there is a flexibility in densities $\rho=\rho(\tilde t)$ that we will get back to below.

Let each particle generated in this way and characterised by a value $\theta_i$ have an uncertainty in the value of said parameter by
\be
P(\theta|i)=\frac{1}{\sigma_i \sqrt{2\pi}}e^{-\frac{(\theta-\theta_i)^2}{2\sigma_i^2}}\,,\quad\left\langle2\sigma_i^2\right\rangle^{-1}=\pi a^2\alpha^2\cosh^{2}(\tilde t/\alpha)
\ee
and let particles interact if they assume the same values of $\theta$ at the same time. So far, the construction produces an interconnected set of particles that, when the time it takes for interaction between particles is rendered in terms of distance, at large scales produces a medium that makes up dS\textsubscript{2}.

Moreover, let the particles interact with the same frequency, be characterised by $\langle \partial_{\tilde t}\theta\rangle=0$ at particle creation (small variations only), and spontaneously and uniformly get {\it less} entangled with time, so that the variance of the position distribution in the particle rest frame at $x_o^\sigma$ obeys
\be
\sigma_i(t_o)=\frac{l_p}{\sqrt{2}}\label{eq.reqB}
\,,\qquad \partial_t\sigma_i(t)\big|_{t=t_o}=0\,,\qquad\frac{\partial_t^2\sigma_i(t)}{\sigma_i(t)}=-\frac{1}{\alpha^2}\,.
\ee
\be\label{eq.reqB2}
\eqref{eq.reqB}\quad\Rightarrow \quad\sigma_i=\frac{l_p}{\sqrt{2}}\cos\left(\frac{t-t_o}{\alpha}\right)\quad \Rightarrow\quad ds^2=\frac{-dt^2+dx^2}{\cos^2([t-t_o]/\alpha)}\,.
\ee
For reasons analogous to the AdS case in \S\ref{s.sAdS}, this causes a propagation of the particles along geodesics in dS\textsubscript{2}. At the particle level, the entanglement between the particles decreases with time through that they spontaneously interact less (with respect to interaction range, not frequency), and in the effective picture that corresponds to an increase in the distance between them.
\\\\
In summary, this quantisation of the spatial dimension in dS\textsubscript{2} gives a set of particles characterised by a parameter $\theta$ taking values on $S^1$, as well as an uncertainty in that same parameter. Interactions are characterised by a Gaussian fall-off with $\Delta\theta$, with a variance that decreases with time, reducing the interaction between the particles. In particular, $\theta\in S^1$ is a marked model difference compared to the energy-related parameter of AdS\textsubscript{2}. It implies a completely different symmetry and dynamics of the dS system.

In addition, in the above quantisation of dS\textsubscript{2} it becomes apparent that an assumption made in the outline of the model ansatz, that of a distribution of basis functions (particles) in the effective spacetime medium $\propto\sqrt{|g|}$ (i.e. an even distribution on the surface embedded in Minkowski space), need not apply to all spacetimes. For AdS, where the spatial configuration is invariant under time evolution and a rescaling of the radial distance, a distribution by $\sqrt{|g|}$ is inferred by symmetry properties\footnote{Shift symmetry in $t$ infers no dependence on $t$. There is also a scale invariance in the metric as expressed in global coordinates (see e.g. the appendix), so there is a symmetry under a rescaling of the radial direction. This infers \mbox{$\rho\propto\sqrt{|g|}$}. From the point of view of the quantised theory, the particle interactions are repulsive, which forces the particles apart with respect to their interaction ranges in an exact correspondence to an even density on the effective spacetime.}, but for dS there is a range of possibly suitable options. Shift symmetry in $\theta$ infers that the density is independent of $\theta$, but no symmetry restricts changes in time. Aside from a distribution by $\sqrt{|g|}$ as in \eqref{eq.rhothe}, a second key scenario to consider is the one in which the number of particles on the $S^1$ described by $\theta$ is kept constant. The model choice of $\rho_\theta$ would depend on what is kept conserved under time evolution, and would have implications for the effective theory. For example, a scenario where the number of particles is kept constant (possibly connecting to unitarity) would at late times result in that the particles no longer interact frequently enough to sustain an interconnected medium. That would cause a phase transition in the effective theory.

Also note that spacetime regions with $R>0$ (expanding spacetime regions) in the present model of quantisation correspond to a reduction in interaction between particles, as if the spacetime simply `floats out' from not being held together by the interactions at a constant level. In such a scenario, an accelerating expansion of the spacetime medium would be a product of the constituent particles increasingly interacting less and less, instead of propelled by an energy. Both scenarios are two sides of the same coin: when something expands it can either do so due to a propelling force, or due to that a force that initially held it together is relaxed.

\subsubsection*{Adding a quantisation of time}
A quantisation of the time in dS\textsubscript{2} can be added to the quantisation of the spatial dimension in a way analogous to how that same quantisation was made for AdS\textsubscript{2} in \S\ref{s.AdS}, except with
\be
\rho_{\tilde t}(\tilde t,\theta)\propto1\,,\quad \tilde t\in \mathbb{R}
\ee
and
\be
\left\langle2\sigma_{\tilde t_i}^2\right\rangle^{-1}=\pi a^2\,,
\ee
as well as the condition that the particles spontaneously become {\it less} entangled with time. Apart from the specifics on {\it(i)} the parameters characterising the particles, {\it (ii)} the probability for a particle with specific values of those parameters to be created, {\it (iii)} the uncertainty in said values and {\it (iv)} how a particle's entanglement with its `environment' evolves, the medium created by the interconnected particles arises in the same way in both cases.

\section{Outlook}\label{s.outlook}
There are several interesting venues in relation to the suggested ansatz for spacetime emergence. The present text has provided examples of what the suggested emergence from entanglement --- in specific, from quantum particles and their interactions --- directly in the spacetime would be characterised by for general $2D$ theories. However, the model is not fully developed. To begin with, it would be of interest to phrase the overlap in position probability distribution of the particles in terms of entanglement entropy, and get a thorough connection to how entanglement entropy connect to emergent spacetime in the gauge/gravity duality. With a better understanding of how the entanglement between the particles can be described, it would also be of interest to compare the gauge/gravity duality results relating to ensembles on the boundary, as observed for Jackiw--Teitelboim gravity \cite{Jackiw:1984je,Teitelboim:1983ux} in \cite{Saad:2019lba}, with the particle entanglement in the suggested ansatz for spacetime emergence. At a first glance, an average over fluctuating particle properties would seem to fit well with a requirement of an ensemble in the boundary theory. The reason for this is simple, and has parallels in fluid dynamics (molecule velocity vs fluid velocity). The particle interactions in the ansatz are not uniquely specified, but include statistical variations around a mean interaction range, which survives into the effective theory. From the perspective of the effective theory, a particle can only have one set interaction range (the mean value), yet to describe a given particle's interaction range as seen from the effective theory, one would have to specify a probability distribution over different ranges. From the point of view of the effective theory, this would appear as a distribution over a set of different effective theories, similar to the concept of an ensemble. In addition, it might be possible (and useful) to phrase the interactions in the suggested ansatz in terms of a matrix model.

Provided that the $2D$ scenarios for spacetime emergence under the suggested ansatz continue to show promise and compatibility with features of spacetime emergence that have been identified, the model should also be extended to higher-dimensional theories, which allow for more degrees of freedom as discussed in \cite{Karlsson:2021ffs}. Constructions involving those need further investigation. In addition, the present analysis focusses solely on spacetime regions where only quantum fluctuations are present. A scenario with out-of-the-vacuum particles passing through would include interactions between the quantum fluctuations and the other particles, altering the properties of the quantum fluctuations and in that way causing backreaction. Precisely how this would occur is outside the scope of this text.

Importantly, note that there is a lot of potential in the suggested ansatz for spacetime emergence. It constitutes a quantisation of spacetime, it is applicable to general spacetimes, and it has characteristics that fit with observed features of AdS spacetimes (emergence from quantum interactions, similarities in distribution to tensor networks, error correcting features etc.). At the same time, the ansatz provides a decidedly novel approach to spacetime emergence, with associated possible new openings to explaining key issues within quantum gravity. For example, a spacetime emergent from quantum interactions between particles directly in the spacetime would constitute an effective theory sensitive to that sufficient interaction actually takes place. At a black hole event horizon, a surface across which equilibrating (two-way) interactions cannot be sustained, this would naturally introduce a surface the effective theory cannot naively be extended across. There, it would (under the suggested ansatz) be reasonable to expect a phase transition from a $D$-dimensional theory to a $(D-1)$-dimensional theory (effectively on the surface of the black hole). Now, such a transition would not necessarily be very different from current scenarios of black hole physics, since the interior of a black hole can be mapped onto the surface of the black hole. The interesting aspect is that the mode of quantisation can provide what so far has been missing altogether: a mechanism for why the black hole event horizon could be a special point in the (effective theory of) spacetime, at all.

There are also possibly interesting aspects of the suggested ansatz that could tie in with dark energy and dark matter. How an accelerating expansion of a spacetime connects to the model was remarked on in \S\ref{s.dS}, and that could provide a new perspective on how one might model the cause of an acceleration. The type of emergence of the effective theory also infers a presence of parallels to how fluids and fluid dynamics originate from particles and their interactions. If one were to regard spacetime not as a rigid background which matter and light passes through, but as a flexible (fluid-like) medium with a movement of its own, certain aspects of dark matter could be rendered in a new light. An intuitive conjecture for the cause of effects related to dark matter would be a simple movement of the spacetime medium itself. Ponder, for example, that when two galaxy clusters collide, the spacetime medium might not equilibrate fast enough to appear like a fixed background, but instead might be left with large-scale differences in density of interaction and relative movement. Such `imprints' (conceptually similar to the imprint caused by a spoon dragged through a fluid with a high viscosity) might give an impression of matter being present, in spite of none (or not enough) being there. Similarly, the spacetime region a galaxy is situated in could possibly (partly) be rotating with the masses in the galaxy itself, in a type of frame-dragging induced to extend across larger distances due to the presence of massive bodies orbiting around the center. 

The above are just a few examples of how a conceptually different spacetime emergence (compared to those currently favoured) might change how spacetime and effects in it are perceived. While the nature of quantum gravity remains elusive, it is important to keep an open mind to what that quantisation actually might entail.

\providecommand{\href}[2]{#2}\begingroup\raggedright\endgroup

\appendix
\section{Embeddings in Minkowski space}\label{s.emb}
Four different reference frames for AdS\textsubscript{2} and dS\textsubscript{2} appear in the main text. For clarity, we here list their embeddings in Minkowski space.

\subsubsection*{AdS\textsubscript{2} coordinates}
In Minkowski space, AdS\textsubscript{2} is obtained through 
\be
ds^2=-dx_1^2-dx_2^2+dx_3^2\,,\quad
-x_1^2-x_2^2+x_3^2=-\alpha^2\,.
\ee
The $(\tilde t,b)$ coordinates in \eqref{eq.AdS2m} are characterised by
\be
\left\{\begin{array}{l}x_1=\sqrt{\alpha^2+b^2}\cos \tilde t\\
x_2=\sqrt{\alpha^2+b^2}\sin \tilde t\\
x_3=b\end{array}\right.\,,\qquad \tilde t\,,\,b\in\mathbb{R}\,,
\ee
and constitute a simple rephrasing of the global coordinates. With $b=\alpha\sinh\rho$ in the above, we get $ds^2=\alpha^2(-\cosh^2\rho\, d\tilde t^2+d\rho^2)$.

The $(t,x)$ coordinates in \eqref{eq.AdS2m} are characterised by
\be
\left[\begin{array}{c}x_1\\x_2\\x_3\end{array}\right]=\frac{\alpha}{\cosh[(t-t_o)/\alpha]}\left[\begin{array}{ccc}\cos (t_o)&-\sin (t_o)&0\\
\sin (t_o)&\cos (t_o)&0\\
0&0&1\end{array}\right]\left[\begin{array}{l} \cosh[x/\alpha]\\ \sinh[(t-t_o)/\alpha]\\\sinh[x/\alpha]\end{array}\right]
\ee
and are constructed so that the reference frame coincides with the rest frame of a particle at any given point $x^\sigma_o$ in the spacetime. At that point, $(t,x)=(t_o,x_o)$. These $(t,x)$ coordinates give a patch of AdS\textsubscript{2} for values close to $(t_o,x_o)$.

\subsubsection*{dS\textsubscript{2} coordinates}
In Minkowski space, dS\textsubscript{2} is obtained through 
\be
ds^2=dx_1^2+dx_2^2-dx_3^2\,,\quad
x_1^2+x_2^2-x_3^2=\alpha^2\,.
\ee
The $(\tilde t,\theta)$ coordinates in \eqref{eq.dS2m} are the global coordinates
\be\label{eq.seeu}
\left\{\begin{array}{l}x_1=\alpha\cosh(\tilde t/\alpha)\cos\theta\\
x_2=\alpha\cosh(\tilde t/\alpha)\sin\theta\\
x_3=\alpha\sinh(\tilde t/\alpha)\end{array}\right.
\ee
where $\tilde t\in \mathbb{R}$ and $\theta$ describes a unit $S^1$.

The $(t,x)$ coordinates in \eqref{eq.dS2m} are characterised by
\begin{gather}\begin{aligned}
\left[\begin{array}{c}x_1\\x_2\end{array}\right]&=\left[\begin{array}{cc}\cos (x_o)&-\sin (x_o)\\
\sin (x_o)&\cos (x_o)\end{array}\right]\left[\begin{array}{c} \frac{r_o\cos[(x-x_o)/\alpha]+r_o\sin[(t-t_o)/\alpha]+\alpha\sin[(x-x_o)/\alpha]}{\cos[(t-t_o)/\alpha]}\\ 
\frac{r_o\sin[(x-x_o)/\alpha]+\alpha\sin[(t-t_o)/\alpha]}{\cos[(t-t_o)/\alpha]\sqrt{r_o^2/\alpha^2-1}}\end{array}\right]\\
x_3&={\rm sgn}(t_o)\sqrt{x_1^2+x_2^2-\alpha^2}\,,\qquad r_o=\sqrt{\alpha^2+t_o^2}\,,\qquad t_o\neq0\,.
\end{aligned}\end{gather}
and for the special case of $x_3(t_o)=0$, which in this parametrisation occurs for $t_o=0$, we have
\be\label{eq.spec}
\left\{\begin{array}{l}x_1=\alpha\cos(x/\alpha)/\cos(t/\alpha)\\
x_2=\alpha\sin(x/\alpha)/\cos(t/\alpha)\\
x_3=\alpha\tan(t/\alpha)\end{array}\right.
\ee
which is a version of the global coordinates. In all, these $(t,x)$ coordinates are constructed so that the reference frame coincides with the rest frame of a particle at any given point $x^\sigma_o$ in the spacetime. At that point, $(t,x)=(t_o,x_o)$. For $t_o\neq0$, these $(t,x)$ coordinates give a patch of dS\textsubscript{2} for values close to $(t_o,x_o)$. The special case of \eqref{eq.spec} covers dS\textsubscript{2}, with $ t\in \mathbb{R}$ and $x$ describing an $S^1$.

\end{document}